\documentclass{article}
\usepackage{ltwol2e}

\input{psfig}
\def\Journal#1#2#3#4{{#1} {\bf #2}, #3 (#4)}

\def\NPB{{\em Nucl. Phys.} B}
\def\PLB{{\em Phys. Lett.}  B}

\def\PRD{{\em Phys. Rev.} D}

\def\be{\begin{equation}}
\def\ee{\end{equation}}
\def\bea{\begin{eqnarray}}
\def\eea{\end{eqnarray}}
\def\slash#1{\setbox0=\hbox{$#1$}#1\hskip-\wd0\dimen0=5pt\advance
\dimen0 by-\ht0\advance\dimen0 by\dp0\lower0.5\dimen0\hbox
to\wd0{\hss\sl/\/\hss}}

\begin{document}

\begin{titlepage}
\null
\vspace{3cm}
\begin{center}
\Large\bf 
A Constituent Quark-Meson Model for Heavy Meson Decays
\end{center}
\vspace{1.5cm}

\begin{center}
Aldo Deandrea\\
\vspace{0.5cm}
Centre de Physique Th\'eorique\footnote{Unit\'e Propre de
Recherche 7061.}, CNRS Luminy, \\
Case 907, F-13288 Marseille Cedex 9, France
\end{center}

\vspace{1.3cm}

\begin{center}
{\bf Abstract}\\[0.5cm]
\parbox{14cm}{I describe a model for heavy meson decays 
based on an effective quark-meson lagrangian. I consider the heavy mesons 
$S$ with spin and parity $J^P=(1^+,0^+), H$ with $J^P=(1^-,0^-)$
and $T^\mu$ with $J^P=(2^+,1^+)$, i.e. S and P wave heavy-light mesons.
The model is constrained by the known symmetries of QCD in the 
$m_Q \to \infty$ for the heavy quarks and chiral symmetry in the light
quark sector.  
Using a very limited number of free parameters it is possible to compute 
several phenomenological quantities, e.g. the leptonic $B$
and $B^{**}$ decay constants; the three universal Isgur-Wise form factors: 
$\xi$, $\tau_{3/2}$, $\tau_{1/2}$, describing the semi-leptonic decays 
$ B \rightarrow D^{(*)} \ell \nu$, $ B \rightarrow D^{**} \ell \nu$; 
the strong and radiative $D^{*}$ decays; the weak
semi-leptonic decays of $B$ and $D$ into light mesons: $\pi,\rho, A_1$.
An overall agreement with data, when available, is achieved.}
\end{center}

\vspace{2cm}

\begin{center}
{\sl To appear in the Proceedings of the\\
XXIXth International Conference on High Energy Physics\\
Vancouver, B.C., Canada, 23--29 July 1998}
\end{center}

\vfil
\noindent
CPT-98/P.3695\\
September 1998

\end{titlepage}

\setcounter{page}{1}

\title{A CONSTITUENT QUARK-MESON MODEL FOR HEAVY MESON DECAYS}

\author{ALDO DEANDREA}

\address{Centre de Physique Th\'eorique, CNRS Luminy, 
Case 907, F-13288 Marseille Cedex 9, France
\\E-mail: deandrea@cpt.univ-mrs.fr}   

\twocolumn[\maketitle\abstracts{I describe a model for heavy meson decays 
based on an effective quark-meson lagrangian. I consider the heavy mesons 
$S$ with spin and parity $J^P=(1^+,0^+), H$ with $J^P=(1^-,0^-)$
and $T^\mu$ with $J^P=(2^+,1^+)$, i.e. S and P wave heavy-light mesons.
The model is constrained by the known symmetries of QCD in the 
$m_Q \to \infty$ for the heavy quarks and chiral symmetry in the light
quark sector.  
Using a very limited number of free parameters it is possible to compute 
several phenomenological quantities, e.g. the leptonic $B$
and $B^{**}$ decay constants; the three universal Isgur-Wise form factors: 
$\xi$, $\tau_{3/2}$, $\tau_{1/2}$, describing the semi-leptonic decays 
$ B \rightarrow D^{(*)} \ell \nu$, $ B \rightarrow D^{**} \ell \nu$; 
the strong and radiative $D^{*}$ decays; the weak
semi-leptonic decays of $B$ and $D$ into light mesons: $\pi,\rho, A_1$.
An overall agreement with data, when available, is achieved.}]

\section{The Model}
The model described in the present paper
is based on an effective constituent quark-meson lagrangian containing
both light and heavy degrees of freedom. It is constrained
by the known symmetries of QCD, i.e. chiral symmetry and heavy quark
symmetry for the heavy quarks in the limit $m_Q\to\infty$ \cite{neurev}. 

The model can be thought of as an intermediate approach between a pure
QCD calculation and an effective theory for heavy and light mesons retaining
only the symmetries of the problem \cite{rep}.

It conjugates the symmetry approach of effective lagrangians with 
well motivated dynamical assumptions on chiral symmetry breaking
and confinement. The effective quark-meson interaction can be for
example deduced 
from partial bosonization of an extended Nambu Jona-Lasinio (NJL) model
\cite{ebert}. In the following this dynamical information will be 
implemented in an effective lagrangian and the few remaining free 
parameters will be fixed by data, thus allowing a number of predictions
based on symmetry and on the implemented dynamics.

This will allow to calculate the parameters of the effective heavy 
meson theory without solving the non-perturbative QCD problem. In
this simplified approach one can hope to describe the essential part of
the QCD behavior, at least in a limited energy range, and extract useful information from it. The model is suitable for the description of
higher spin heavy mesons as they can be included in the formalism
in a very easy way \cite{noi} (see also \cite{feld}). On the contrary the 
inclusion of higher order corrections, albeit possible, requires
the determination of new free parameters, which proliferate
as new orders are added to the expansion. In this sense the model
allows a simple and intuitive approach to heavy-meson processes
if it is kept at lowest order, while it loses part of its predictive power
if corrections have to be included.

\subsection{Heavy meson field}\label{subsec:hmf}
In order to implement the heavy quark symmetries in the spectrum of
physical states the wave function of a heavy meson has to be independent
of the heavy quark flavor and spin. It can be characterized
by the total angular momentum $s_\ell$ of the light degrees of freedom.
To each value of $s_\ell$ corresponds
a degenerate doublet of states with angular momentum $J=s_\ell \pm 1/2$.
The mesons $P$ and $P^*$ form the spin-symmetry
doublet corresponding to $s_\ell= 1/2$ (for charm for instance,
they correspond to $D$ and $D^*$).

The negative parity spin doublet $(P,P^*)$ can be represented by a
$4 \times 4$ Dirac matrix $H$, with one spinor index for the heavy quark
and the other for the light degrees of freedom.

An explicit matrix representation is:
\begin{eqnarray}
H &=& \frac{(1+\slash v)}{2}\; [P_{\mu}^*\gamma^\mu - P \gamma_5 ]\\
{\bar H} &=& \gamma_0 H^{\dagger} \gamma_0~~.
\end{eqnarray}
Here $v$ is the heavy meson velocity, $v^\mu P^*_{a\mu}=0$ and
$M_H=M_P=M_{P^*}$. Moreover $\slash v H=-H \slash v =H$, ${\bar H}
\slash v=-\slash v {\bar H}={\bar H}$ and $P^{*\mu}$ and $P$ are annihilation
operators normalized as follows:
\begin{eqnarray}
\langle 0|P| Q{\bar q} (0^-)\rangle & =&\sqrt{M_H}\\
\langle 0|{P^*}^\mu| Q{\bar q} (1^-)\rangle & = & \epsilon^{\mu}\sqrt{M_H}~~.
\end{eqnarray}
The formalism for higher spin states was introduced by Falk and 
Luke \cite{falk}.
I shall consider only the $S$ and $P$-waves of the system $Q\bar q$.
The heavy quark effective theory predicts two distinct multiplets, one
containing a $0^+$ and a $1^+$ degenerate state, and the other one a $1^+$ and
a $2^+$ state. In matrix notation, analogous to the one used for the
negative parity states, they are described by
\begin{equation}
S={{1+\slash v}\over 2}[P_{1\mu}^{*\prime} \gamma^\mu\gamma_5-P_0]
\end{equation}
and
\begin{equation}
T^\mu={\frac {1+\slash v}{2}}\left[P_2^{* \mu\nu}\gamma_\nu-\sqrt{\frac 3
2}
P^*_{1\nu}\gamma_5\left(g^{\mu\nu}-\frac 1 3
\gamma^\nu(\gamma^\mu-v^\mu)\right)
\right].
\end{equation}
These two multiplets have $s_\ell=1/2$ and $s_\ell=3/2$ respectively, where
$s_\ell$ is conserved together with the spin $s_Q$ in the infinite quark
mass limit because $\vec J={\vec s}_\ell+{\vec s}_Q$.
\vspace*{-1.8pt}   %optional

\subsection{Meson-Quark Interaction}\label{subsec:mqi}
The light degrees of freedom,
i.e. the light quark fields $\chi$ and the pseudo-scalar
$SU(3)$ octet of mesons $\pi$ are introduced using the chiral lagrangian:
\begin{eqnarray}
{\cal L}_{\ell \ell}&=&{\bar \chi} (i  D^\mu \gamma_\mu +g_A
{\cal A}^\mu \gamma_\mu \gamma_5) \chi - m {\bar \chi}\chi \nonumber\\
&+& {f_{\pi}^2\over 8} \partial_{\mu} \Sigma^{\dagger} \partial^{\mu}
\Sigma .
\end{eqnarray}
Here $D_\mu = \partial_\mu-i {\cal V}_\mu$,
$\xi=\exp(i\pi/f_\pi )$, $\Sigma=\xi^2$, $f_\pi=130$ MeV and
\begin{eqnarray}
{\cal V}^\mu &=& {1\over 2} (\xi^\dagger \partial^\mu \xi +\xi \partial^\mu
\xi^\dagger) \nonumber \\
{\cal A}^\mu &=& {i\over 2} (\xi^\dagger \partial^\mu \xi -\xi \partial^\mu
\xi^\dagger)~. \label{av}
\end{eqnarray}
The term with $g_A$ is the coupling of pions to light quarks; it will not
be used in the sequel. It is a free parameter, but in NJL model $g_A=1$.

One can introduce a quark-meson effective lagrangian
involving heavy and light quarks and heavy mesons. At lowest order one has:
\begin{eqnarray}
{\cal L}_{h \ell}&=&{\bar Q}_v i v\cdot \partial Q_v
-\left( {\bar \chi}({\bar H}+{\bar S}+ i{\bar T}_\mu
{D^\mu \over {\Lambda_\chi}})Q_v +h.c.\right)\nonumber \\
&+&\frac{1}{2 G_3} {\mathrm {Tr}}[({\bar H}+{\bar S})(H-S)]
+\frac{1}{2 G_4}
{\mathrm {Tr}} [{\bar T}_\mu T^\mu ] \label{qh1}
\end{eqnarray}
where the meson fields $H,~S,~T$ have been defined is section \ref{subsec:hmf}, 
$Q_v$ is the effective
heavy quark field, $G_3$, $G_4$ are coupling constants
and $\Lambda_\chi$ ($= 1$ GeV) has been introduced for dimensional reasons. 
Lagrangian (\ref{qh1}) has heavy spin and flavor symmetry. 
This lagrangian comprises three
terms containing respectively $H$, $S$ and $T$.
Note that the fields $H$ and $S$ have the same coupling constant.
In doing this one assumes that this effective quark-meson
lagrangian can be justified as a remnant of a four quark interaction of
the NJL type by partial bosonization. 
\vspace*{-1.8pt}   %optional

\subsection{Cut-Off Prescription}\label{subsec:cop}
The cut-off prescription is the way in which part of the dynamical 
information regarding QCD is introduced in the model, this is why it 
is crucial and is part of the definition of the model. 
As the heavy mesons are described consistently with HQET,
the heavy quark propagator in the loop contains the
residual momentum $k$ which arises from the interaction with the light
degrees of freedom. It is natural to assume an ultraviolet cut-off
on the loop momentum of the order of $\Lambda \simeq 1$ GeV, even
if the heavy quark mass is larger than the cut-off.

In the infrared the model is not confining and 
its range of validity can not be extended below energies of the order of
$\Lambda_{QCD}$. In practice one introduces an infrared cut-off $\mu$, 
to take this into account.

The cut-off prescription is implemented via a proper time regularization 
(a different choice is followed in \cite{bardeen}). After continuation to
the Euclidean space it reads, for the light quark propagator:
\begin{equation}
\int d^4 k_E \frac{1}{k_E^2+m^2} \to \int d^4 k_E \int^{1/
\Lambda^2}_{1/\mu^2} ds\; e^{-s (k_E^2+m^2)}\label{cutoff}
\end{equation}
where $\mu$ and $\Lambda$ are infrared and ultraviolet cut-offs.

Reasonable values are $\Lambda\simeq 1$ GeV, $m\simeq \mu \simeq
10^2$ MeV. The cut-off prescription is similar to the
one in \cite{ebert}, with $\Lambda=1.25$ GeV; 
the numerical results are not strongly dependent on the value of $\Lambda$. 
The constituent mass $m$ in the NJL models represents the order
parameter discriminating between the phases of broken and unbroken chiral
symmetry and can be fixed by solving a gap equation, which gives $m$ as a
function of the scale mass $\mu$ for given values of the other parameters. In
the second paper of Ebert et al. \cite{ebert} 
the values $m=300$ MeV and $\mu=300$ MeV are used
and we shall assume the same values. As shown there, for smaller
values of $\mu$, $m$ is constant (=300 MeV) while for
much larger values of $\mu$, it  decreases and in
particular it vanishes for $\mu=550$ MeV.

\section{Analytical and Numerical Results}
\subsection{Decay constants}\label{subsec:dcc}
The leptonic decay constants ${\hat F}$ and ${\hat F}^+$
are defined as follows:
\begin{eqnarray}
\langle 0|{\bar q}
\gamma^\mu \gamma_5 Q  |H( 0^-, v)\rangle  & =&i \sqrt{M_H} v^\mu
{\hat F}\\
\langle 0|{\bar q}
\gamma^\mu Q |S( 0^+, v)\rangle & =&i \sqrt{M_S} v^\mu {\hat F}^+~~.
\end{eqnarray}
and they can be computed by a loop calculation, where the heavy meson 
interacts with the heavy and light quarks (via the interaction introduced 
in (\ref{qh1}) and then those interact with the current.
The result is
\begin{eqnarray}
{\hat F}&=&\frac{\sqrt{Z_H}}{G_3}\\
{\hat F}^+ &=& \frac{\sqrt{Z_S}}{G_3}
\end{eqnarray}
where $Z_H$ and $Z_S$ are the field renormalization constants. 
Detailed results can be found in Deandrea et al. \cite{noi}. Values for the 
renormalization constants and couplings can be read in Table \ref{tab:1tab}
for three values of the parameter $\Delta_H$.
\begin{table}
\begin{center}
\caption{Renormalization constants and couplings.
$\Delta_H$ in GeV; $G_3$, $G_4$ in GeV$^{-2}$,
$Z_j$ in GeV$^{-1}$.}\label{tab:1tab}
\vspace{0.2cm}
\begin{tabular}{|c|c|c|c|c|c|} 
\hline 
\raisebox{0pt}[12pt][6pt]{$\Delta_H$} & 
\raisebox{0pt}[12pt][6pt]{$1/G_3$} & 
\raisebox{0pt}[12pt][6pt]{$Z_H$} &
\raisebox{0pt}[12pt][6pt]{$Z_S$} &
\raisebox{0pt}[12pt][6pt]{$Z_T$} &
\raisebox{0pt}[12pt][6pt]{$1/G_4$} \\
\hline
\hline
\raisebox{0pt}[12pt][6pt]{0.3} & 
\raisebox{0pt}[12pt][6pt]{0.16} & 
\raisebox{0pt}[12pt][6pt]{4.17} & 
\raisebox{0pt}[12pt][6pt]{1.84} &
\raisebox{0pt}[12pt][6pt]{2.95} &
\raisebox{0pt}[12pt][6pt]{0.15} \\
\hline
\raisebox{0pt}[12pt][6pt]{0.4} & 
\raisebox{0pt}[12pt][6pt]{0.22} & 
\raisebox{0pt}[12pt][6pt]{2.36} & 
\raisebox{0pt}[12pt][6pt]{1.14} &
\raisebox{0pt}[12pt][6pt]{1.07} &
\raisebox{0pt}[12pt][6pt]{0.26} \\
\hline
\raisebox{0pt}[12pt][6pt]{0.5} & 
\raisebox{0pt}[12pt][6pt]{0.345} & 
\raisebox{0pt}[12pt][6pt]{1.14} & 
\raisebox{0pt}[12pt][6pt]{0.63} &
\raisebox{0pt}[12pt][6pt]{0.27} &
\raisebox{0pt}[12pt][6pt]{0.66} \\
\hline
\end{tabular}
\end{center}
\end{table}
\vspace*{3pt}
The numerical results for the decay constants can be found in Table 
\ref{tab:2tab}.
\begin{table}
\begin{center}
\caption{${\hat F}$ and ${\hat F}^+$ for
various values of $\Delta_H $. $\Delta_H$ in GeV, leptonic constants
in GeV$^{3/2}$.}\label{tab:2tab}
\vspace{0.2cm}
\begin{tabular}{|c|c|c|} 
\hline 
\raisebox{0pt}[12pt][6pt]{$\Delta_H$} & 
\raisebox{0pt}[12pt][6pt]{$\hat F$} & 
\raisebox{0pt}[12pt][6pt]{${\hat F}^+$} \\
\hline
\hline
\raisebox{0pt}[12pt][6pt]{0.3} & 
\raisebox{0pt}[12pt][6pt]{0.33} & 
\raisebox{0pt}[12pt][6pt]{0.22} \\
\hline
\raisebox{0pt}[12pt][6pt]{0.4} & 
\raisebox{0pt}[12pt][6pt]{0.34} & 
\raisebox{0pt}[12pt][6pt]{0.24} \\
\hline
\raisebox{0pt}[12pt][6pt]{0.5} & 
\raisebox{0pt}[12pt][6pt]{0.37} & 
\raisebox{0pt}[12pt][6pt]{0.27} \\
\hline
\end{tabular}
\end{center}
\end{table}
\vspace*{3pt}
Neglecting logarithmic corrections, ${\hat F}$ and ${\hat F}^+$ are related,
in the infinite heavy quark mass limit, to the leptonic decay constant 
$f_B$ and $f^+$. For example, for $\Delta_H=400$ MeV, one obtains from 
Table \ref{tab:2tab}:
\begin{eqnarray}
f_B &\simeq& 150 \;{\mathrm {MeV}}\\
f^+ &\simeq& 100 \;{\mathrm {MeV}}\;.
\end{eqnarray}
\vspace*{-1.8pt}   %optional

\subsection{Semi-leptonic Decays and Form Factors}\label{subsec:sdff}
As an example of the quantities that can be analytically calculated
in the model, one can examine the Isgur-Wise function $\xi$:
\begin{eqnarray}
\langle D(v^\prime)&|&{\bar c} \gamma_\mu (1-\gamma_5) b|
B(v)\rangle = \sqrt{M_B M_D} \nonumber \\
&&\times C_{cb} \xi(\omega) (v_\mu + v^{\prime }_{\mu})
\end{eqnarray}
\begin{eqnarray}
&&\langle D^*(v^\prime, \epsilon)|{\bar c} \gamma_\mu (1- \gamma_5) b| B(v)
\rangle = \sqrt{M_B M_{D^*}} C_{cb}\; \xi(\omega)\nonumber \\
&&\times [i \epsilon_{\mu \nu \alpha \beta}
\epsilon^{*\nu}v^{\prime \alpha}v^\beta 
- (1+\omega)\epsilon^*_\mu+ (\epsilon^*\cdot v)v^\prime_\mu]
\end{eqnarray}
where $\omega= v \cdot v^\prime$ and $C_{cb}$
is a coefficient containing
logarithmic corrections depending on $\alpha_s$; within our approximation
it can be put equal to 1: $C_{cb}=1$. We also note that,
in the leading order we are considering here $\xi(1)=1$.

One finds \cite{ebert}:
\begin{equation}
\xi(\omega)=Z_H \left[ \frac{2}{1+\omega} I_3(\Delta_H)+\left( m+\frac{2
\Delta_H}{1+\omega} \right)
I_5(\Delta_H, \Delta_H,\omega) \right] ~~.\label{xi}
\end{equation}
\begin{figure}
\center
\psfig{figure=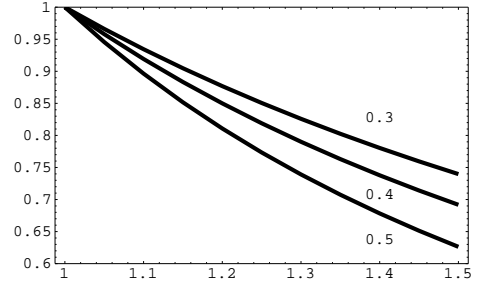,height=1.5in}
\caption{Isgur-Wise form factor at different $\Delta$ values.}
\label{fig:iw}
\end{figure}
The $\xi$ function is plotted in Fig. \ref{fig:iw}.

The integrals $I_3,~I_5$ can be found in the Appendix.

One can compute in a similar way the form factors describing 
the semi-leptonic
decays of a meson belonging to the fundamental negative
parity multiplet $H$ into the
positive parity mesons in the $S$ and $T$ multiplets.
Examples of these decays are
\begin{equation}
B \rightarrow D^{**} \ell \nu \label{bd**}
\end{equation}
where $D^{**}$ can be either a $S$ state
(i.e. a $0^+$ or $ 1^+$ charmed meson having
$s_{\ell}=1/2$)
or  a $T$ state ( i.e. a $2^+$ or $ 1^+$ charmed meson having $s_{\ell}=3/2$).

The decays in (\ref{bd**}) are described by two form
factors $\tau_{1/2}, \tau_{3/2}$ \cite{IW2} which
can be computed by a loop calculation similar to the one used to obtain
$\xi(\omega)$. The result is
\begin{eqnarray}
\tau_{1/2}(\omega)&=&\frac{\sqrt{Z_H Z_S}}{2(1-\omega)}
\Big[ I_3(\Delta_S)-I_3(\Delta_H) \nonumber \\
&+&\left(\Delta_H - \Delta_S + m(1-\omega) \right)
I_5(\Delta_H, \Delta_S,\omega) \Big]
\end{eqnarray}
and

\begin{eqnarray}
&\tau_{3/2}&(\omega)=-{{\sqrt{Z_H\,Z_T}}\over{{\sqrt{3}}}} \nonumber \\
&\times &\Big[m \Big( {{I_3(\Delta_H) - I_3(\Delta_T) -
              \left( \Delta_H - \Delta_T \right) \,
               I_5(\Delta_H,\Delta_T,\omega )}\over
            {2\,\left( 1 - \omega  \right) }} \nonumber \\
&-&{{I_3(\Delta_H) + I_3(\Delta_T)  +
              \left( \Delta_H + \Delta_T \right)  \,
               I_5(\Delta_H,\Delta_T,\omega )}\over
            {2\,\left( 1 + \omega  \right) }} \Big) \nonumber\\
&-& {{1}\over {2\,\left( -1 - \omega  + {{\omega }^2} +  {{\omega }^3}
\right) }} \Big( -3\,S(\Delta_H,\Delta_T, \omega ) \nonumber \\
&-&\left( 1 - 2\,\omega   \right) \,S(\Delta_T,\Delta_H,
\omega ) + (1-{{\omega }^2}) \,T(\Delta_H,\Delta_T,\omega  ) \nonumber \\
&-& 2\, (1 - 2\,\omega ) \,U(\Delta_H,\Delta_T,\omega ) \Big)  \Big]
\end{eqnarray}
where the integrals $S,T,U$ are defined in the Appendix.

The numerical results are reported in Table \ref{tab:3tab}. For a 
comparison with other calculations of these form factors see \cite{pe}. 
\begin{table}
\begin{center}
\caption{Form factors and slopes. $\Delta_H$ in GeV.}\label{tab:3tab}
\vspace{0.2cm}
\begin{tabular}{|c|c|c|c|c|c|c|} 
\hline
\raisebox{0pt}[12pt][6pt]{$\Delta_H$} & 
\raisebox{0pt}[12pt][6pt]{$\xi(1)$} & 
\raisebox{0pt}[12pt][6pt]{$\rho^2_{IW}$} &
\raisebox{0pt}[12pt][6pt]{$\tau_{1/2}(1)$} &
\raisebox{0pt}[12pt][6pt]{$\rho^2_{1/2}$} &
\raisebox{0pt}[12pt][6pt]{$\tau_{3/2}(1)$} &
\raisebox{0pt}[12pt][6pt]{$\rho^2_{3/2}$} \\
\hline
\hline
\raisebox{0pt}[12pt][6pt]{0.3} & 
\raisebox{0pt}[12pt][6pt]{1} & 
\raisebox{0pt}[12pt][6pt]{0.72} & 
\raisebox{0pt}[12pt][6pt]{0.08} & 
\raisebox{0pt}[12pt][6pt]{0.8} &
\raisebox{0pt}[12pt][6pt]{0.48} &
\raisebox{0pt}[12pt][6pt]{1.4} \\
\hline
\raisebox{0pt}[12pt][6pt]{0.4} & 
\raisebox{0pt}[12pt][6pt]{1} & 
\raisebox{0pt}[12pt][6pt]{0.87} & 
\raisebox{0pt}[12pt][6pt]{0.09} & 
\raisebox{0pt}[12pt][6pt]{1.1} &
\raisebox{0pt}[12pt][6pt]{0.56} &
\raisebox{0pt}[12pt][6pt]{2.3} \\
\hline
\raisebox{0pt}[12pt][6pt]{0.5} & 
\raisebox{0pt}[12pt][6pt]{1} & 
\raisebox{0pt}[12pt][6pt]{1.14} & 
\raisebox{0pt}[12pt][6pt]{0.09} & 
\raisebox{0pt}[12pt][6pt]{2.7} &
\raisebox{0pt}[12pt][6pt]{0.67} &
\raisebox{0pt}[12pt][6pt]{3.0} \\
\hline
\end{tabular}
\end{center}
\end{table}
\vspace*{3pt}

An important test of our approach is represented by the
Bjorken sum rule, which states
\begin{equation}
\rho^2_{IW}=\frac{1}{4}+\sum_k \left[|\tau_{1/2}^{(k)}(1)|^2~+~
2|\tau_{3/2}^{(k)}(1)|^2\right]~.
\end{equation}
Numerically we find that the first excited resonances, i.e. the
$S$ and $T$ states ($k=0$) practically saturate
the sum rule for all the three values of $\Delta_H$.

\subsection{Strong decays}
The model can be used to calculate strong coupling constants, such as
those concerning the decays:
\begin{eqnarray}
&H& \rightarrow H \pi \label{HHpi}\\
&S& \rightarrow H\pi \label{SHpi}
\end{eqnarray}
The constant
$g_{D^{*}D\pi}$ is related to the strong coupling constant
of the effective meson field theory $g$ appearing in the heavy
meson effective lagrangian \cite{rep}
\begin{equation}
{\cal L}=ig {\mathrm {Tr}}(\overline{H} H \gamma^{\mu}\gamma_5{\cal A}_{\mu})
\; +\; \left[ih\; {\mathrm {Tr}}(\overline{H} S \gamma^{\mu}\gamma_5
{\cal A}_{\mu})  \; + \; {\mathrm {h.c.}}\right]
\label{lg}
\end{equation}
by the relation
\begin{equation}
g_{D^{*}D\pi}=\frac{2m_{D}}{f_\pi}g
\end{equation}
valid in the $m_Q \to \infty$ limit. 

Numerically one gets
\begin{equation}
g=0.456\pm0.040
\end{equation}
where the central value corresponds to $\Delta_H=0.4$ GeV and the
lower (resp. higher) value corresponds to $\Delta_H=0.3$ GeV
(resp. $\Delta_H=0.5$ GeV). In an analogous way one obtains
\begin{equation}
h=-0.85\pm0.02
\end{equation}
The details of the calculation can be found in Deandrea et al. \cite{noi}.
Once the coupling constants are calculated, it is possible to make 
predictions for branching ratios in strong heavy mesons decays. The results
are given in Table \ref{tab:4tab} and are in good agreement with experimental 
data.

\begin{table}
\begin{center}
\caption{Theoretical and  experimental $D^*$ branching ratios (\%).
Theoretical values are computed with $\Delta_H=0.4$ GeV.}\label{tab:4tab}
\vspace{0.2cm}
\begin{tabular}{|c|c|c|} 
\hline 
\raisebox{0pt}[12pt][6pt]{Decay mode} & 
\raisebox{0pt}[12pt][6pt]{Br (\%)} & 
\raisebox{0pt}[12pt][6pt]{Exp} \\
\hline
\hline
\raisebox{0pt}[12pt][6pt]{${D^*}^0\to D^0 \pi^0$} & 
\raisebox{0pt}[12pt][6pt]{65.5} & 
\raisebox{0pt}[12pt][6pt]{$61.9 \pm 2.9$} \\
\hline
\raisebox{0pt}[12pt][6pt]{${D^*}^+\to D^0 \pi^+$} & 
\raisebox{0pt}[12pt][6pt]{71.6} & 
\raisebox{0pt}[12pt][6pt]{$68.3\pm 1.4$} \\
\hline
\raisebox{0pt}[12pt][6pt]{${D^*}^+\to D^+ \pi^0$} & 
\raisebox{0pt}[12pt][6pt]{28.0} & 
\raisebox{0pt}[12pt][6pt]{$30.6 \pm 2.5$} \\
\hline
\end{tabular}
\end{center}
\end{table}

\section{Conclusions}

Starting from an effective lagrangian at the level of mesons and constituent
quarks, one can calculate meson transition amplitudes by evaluating loops 
of heavy and light quarks. In this way it is possible to compute the 
Isgur-Wise function, the form factors $\tau_{1/2}$ and $\tau_{3/2}$, 
the leptonic decay 
constant ${\hat F}$ and ${\hat F}^+$, and many other quantities, such as 
radiative and strong couplings and decays which are only briefly 
mentioned in this short note (see for details \cite{noi} and \cite{new}). 
The agreement with data, when available, is very good in most cases. 
The model is able to describe a number of essential features of 
heavy meson physics in a simple and compact way.

\section*{Acknowledgments}
I acknowledge the support of a ``Marie Curie'' TMR research fellowship
of the European Commission under contract ERBFMBICT960965. The subject
described in this paper is based on work in collaboration with
R. Gatto, G. Nardulli, A. Polosa and in the early stage of the work 
with N. Di Bartolomeo. I wish to thank them all.
Centre de Physique Th\'eorique is Unit\'e Propre de
Recherche 7061.

\section*{Appendix}

The integrals used in the paper are listed in this appendix. A
more exhaustive list of integrals with proper time regularization
useful for calculations in the model described here can be found in
the appendix of \cite{noi}.

\begin{eqnarray}
I_1&=&\frac{iN_c}{16\pi^4} \int^{reg} \frac{d^4k}{(k^2 - m^2)}
={{N_c m^2}\over {16 \pi^2}} \Gamma(-1,{{{m^2}}
\over {{{\Lambda}^2}}},{{{m^2}}\over {{{\mu }^2}}}) \nonumber
\\
I_3(\Delta) &=& - \frac{iN_c}{16\pi^4} \int^{\mathrm {reg}}
\frac{d^4k}{(k^2-m^2)(v\cdot k + \Delta + i\epsilon)}\nonumber \\
&=&{N_c \over {16\,{{\pi }^{{3/2}}}}}
\int_{1/{{\Lambda}^2}}^{1/{{\mu }^2}} {ds \over {s^{3/2}}}
\; e^{- s( {m^2} - {{\Delta }^2} ) }\;
\left( 1 + {\mathrm {erf}} (\Delta\sqrt{s}) \right) \nonumber
\end{eqnarray}
where $m$ is the constituent light quark mass of the order of 300 MeV
and
\begin{eqnarray}
\Gamma(\alpha,x_0,x_1) &=& \int_{x_0}^{x_1} dt\;  e^{-t}\; t^{\alpha-1}
\nonumber
\end{eqnarray}
is the generalized incomplete gamma function and erf is the error function.
In order to keep $I_5$ and $I_6$ in a short form one can introduce the
function:
\begin{eqnarray}
&&\sigma(x,\Delta_1,\Delta_2,\omega)={{{\Delta_1}\,\left( 1 - x \right)  +
{\Delta_2}\,x}\over {{\sqrt{1 + 2\,\left(\omega -1  \right) \,x +
2\,\left(1-\omega\right) \,{x^2}}}}} \nonumber
\end{eqnarray}
Finally one has:
\begin{eqnarray}
I_5&&(\Delta_1,\Delta_2,\omega) = \int_{0}^{1} dx \frac{1}{1+2x^2 (1-\omega)+2x
(\omega-1)}\times\nonumber\\
&&\Big[ \frac{6}{16\pi^{3/2}}\int_{1/\Lambda^2}^{1/\mu^2} ds~\sigma
\; e^{-s(m^2-\sigma^2)} \; s^{-1/2}\; (1+ {\mathrm {erf}}
(\sigma\sqrt{s})) +\nonumber\\
&&\frac{6}{16\pi^2}\int_{1/\Lambda^2}^{1/\mu^2}
ds \; e^{-s(m^2-2\sigma^2)}\; s^{-1}\Big] \nonumber
\end{eqnarray}

\begin{eqnarray}
I_6&&(\Delta_1,\Delta_2,\omega) = I_1 \int_{0}^{1} dx
\frac{\sigma}{1+2x^2(1-\omega) + 2x(\omega-1)}\nonumber\\
&-&\frac{N_c}{16 \pi^{3/2}}\int_{0}^{1} dx
\frac{1}{1+2x^2(1-\omega) + 2x(\omega-1)}\times\nonumber\\
&&\int_{1/\Lambda^2}^{1/\mu^2} \frac{ds}{s^{3/2}}
e^{-s(m^2-\sigma^2)}\Big\{\sigma
[1+{\mathrm {erf}}(\sigma\sqrt{s})] \times \nonumber \\
&&[1+2s(m^2-\sigma^2)]
+2 {\sqrt{\frac{s}{\pi}}} e^{-s\sigma^2}\left[
\frac{3}{2s} + (m^2-\sigma^2)\right] \nonumber
\end{eqnarray}

\begin{eqnarray}
S(\Delta_1,\Delta_2,\omega)&=&\Delta_1\,I_3(\Delta_2) + \omega \,
\left( I_1 + \Delta_2 \,I_3(\Delta_2 ) \right)  \nonumber \\
&+&{{\Delta_1}^2}\,I_5(\Delta_1,\Delta_2 ,\omega ) \nonumber \\
T(\Delta_1,\Delta_2,\omega)&=&{m^2}\,I_5(\Delta_1,\Delta_2 ,\omega ) +
I_6(\Delta_1,\Delta_2 ,\omega ) \nonumber \\
U(\Delta_1,\Delta_2,\omega)&=&I_1 + \Delta_2 \,I_3(\Delta_2 ) + \Delta_1\,
I_3(\Delta_1) \nonumber \\
&+& \Delta_2 \,\Delta_1\,I_5(\Delta_1,\Delta_2 ,\omega )
\end{eqnarray}


\section*{References}
\begin{thebibliography}{99}

\bibitem{neurev}
For further references see the review papers:
H. Georgi, contribution to the {\it Proceedings of TASI 91}, R.K. Ellis ed.,
World Scientific, Singapore,1991;
B. Grinstein, contribution to {\it High Energy Phenomenology}, R. Huerta and
M.A. Peres eds., World Scientific, Singapore, 1991;
N. Isgur and M. Wise, contribution to {\it Heavy Flavours}, A. Buras and M.
Lindner eds., World Scientific, Singapore,1992;
M. Neubert, {\it Phys. Rep.} {\bf 245} 259 (1994).

\bibitem{rep}
R. Casalbuoni et al., {\it Phys. Rep.} {\bf 281} 145 (1997).

\bibitem{ebert}
D. Ebert, T. Feldmann, R. Friedrich and H. Reinhardt,
\Journal{\NPB}{434}{619}{1995}; D. Ebert, T.~Feldmann and H.~Reinhardt,
\Journal{\PLB}{388}{154}{1996}.

\bibitem{noi}
A. Deandrea, N. Di Bartolomeo, R. Gatto, G. Nardulli, A.D. Polosa,
\Journal{\PRD}{58}{034004}{1998}.

\bibitem{feld}
T. Feldmann, hep-ph/9606451 and also PhD Thesis (in German) available
at http://wptu38.physik.uni-wuppertal.de/ $\sim$feldmann/pub/thesis.ps

\bibitem{falk}
A. Falk and M. Luke \Journal{\PLB}{292}{119}{1992}.

\bibitem{bardeen}
W. H. Bardeen and C. T. Hill, \Journal{\PRD}{49}{409}{1994}.
B. Holdom and M. Sutherland, \Journal{\PRD}{47}{5067}{1993},
B. Holdom and M. Sutherland, \Journal{\PLB}{313}{447}{1993},
B. Holdom and M. Sutherland, \Journal{\PRD}{48}{5196}{1993},
B. Holdom, M. Sutherland and J. Mureika, \Journal{\PRD}{49}{2359}{1994},
M. Sutherland, B. Holdom, S. Jaimungal, and R. Lewis, \Journal{\PRD} 
{51}{5053}{1995}.

\bibitem{IW2}
N. Isgur and M. B. Wise, \Journal{\PRD}{43}{819}{1991}.

\bibitem{pe}
V. Morenas et al., \Journal{\PRD}{56}{5668}{1997}.

\bibitem{new}
A. Deandrea, R. Gatto, G. Nardulli, A.D. Polosa, BARI-TH/98-314

\end{thebibliography}
\end{document}